\documentclass[aps,prb,twocolumn,superscriptaddress]{revtex4-2}
\usepackage[T1]{fontenc}

\usepackage{bm}
\renewcommand{\vec}[1]{\bm{#1}}

\newcommand{\h}{\hbar}
\newcommand{\kb}{k_\mathrm{B}}
\newcommand{\kbt}{\kb T}

\newcommand{\hc}{\mathrm{H.c.}}

\renewcommand{\a}{\alpha}

\newcommand{\z}{\zeta}
\renewcommand{\t}{\tau}

\renewcommand{\l}{\lambda}

\renewcommand{\r}{\rho}

\renewcommand{\o}{\omega}

\renewcommand{\O}{\Omega}

\renewcommand{\i}{\mathrm{i}} %Imaginary unit

\newcommand{\omc}{\omega_\mathrm{c}}

\newcommand{\bj}{b_j^{\mathstrut}}
\newcommand{\bcj}{b^\dagger_j}

\newcommand{\bi}{B}
\newcommand{\gr}{G}
\newcommand{\xh}{X_{\mathrm{H}}}
\newcommand{\xv}{X_{\mathrm{V}}}
\newcommand{\xj}{X_j}

\newcommand{\levels}{\bi, \gr, \xh, \xv}

\newcommand{\pol}{\mathrm{H}, \mathrm{V}}

\newcommand{\pbxj}{\dyad{\bi}{\xj}}

\newcommand{\pxjg}{\dyad{\xj}{\gr}}

\newcommand{\Oj}{\O_j}

\newcommand{\oxj}{\omega_{\xj}}

\newcommand{\og}{\omega_{\gr}}

\newcommand{\vq}{{\vec{q}}}
\newcommand{\oq}{{\o_\vq}}
\newcommand{\aq}{a_{\vq}}
\newcommand{\acq}{a_{\vq}^\dagger}

\newcommand{\Hqdc}{H_\mathrm{QD-Cav}}
\newcommand{\Hqdl}{H_\mathrm{QD-L}}
\newcommand{\Hqdp}{H_\mathrm{QD-Ph}}

\usepackage{graphicx}
\usepackage{xcolor}
\usepackage{mathtools}
\usepackage{physics}
\usepackage{cleveref}
\usepackage{units}
\usepackage{siunitx}
\usepackage{dcolumn}% Align table columns on decimal point

\begin{document}

%\preprint{}

\title{Pulse shaping for on-demand emission of single Raman photons \\ from a quantum-dot biexciton}


\author{Tom Praschan}
%\thanks{}
\affiliation{Department of Physics and Center for Optoelectronics and Photonics Paderborn (CeOPP), Paderborn University, Warburger Strasse 100, 33098 Paderborn, Germany}

\author{Dirk Heinze}
\affiliation{Department of Physics and Center for Optoelectronics and Photonics Paderborn (CeOPP), Paderborn University, Warburger Strasse 100, 33098 Paderborn, Germany}

\author{Dominik Breddermann}
\affiliation{Department of Physics and Center for Optoelectronics and Photonics Paderborn (CeOPP), Paderborn University, Warburger Strasse 100, 33098 Paderborn, Germany}

\author{Artur Zrenner}
\affiliation{Department of Physics and Center for Optoelectronics and Photonics Paderborn (CeOPP), Paderborn University, Warburger Strasse 100, 33098 Paderborn, Germany}

\author{Andrea Walther}
\affiliation{Department of Mathematics, Humboldt-Universit\"at zu Berlin, Unter den Linden 6, 10099 Berlin, Germany}

\author{Stefan Schumacher}
%\thanks{}
\affiliation{Department of Physics and Center for Optoelectronics and Photonics Paderborn (CeOPP), Paderborn University, Warburger Strasse 100, 33098 Paderborn, Germany}
\affiliation{Wyant College of Optical Sciences, University of Arizona, Tucson, Arizona 85721, USA}


\date{\today}

\begin{abstract}

Semiconductor quantum dots embedded in optical cavities are promising on-demand sources of single photons. Here, we theoretically study single photon emission from an optically driven two-photon Raman transition between the biexciton and the ground state of a quantum dot. The advantage of this process is that it allows all-optical control of the properties of the emitted single photon with a laser pulse. However, with the presence of other decay channels and excitation-induced quantum interference, on-demand emission of the single Raman photon is generally difficult to achieve. Here we show that laser pulses with non-trivial shapes can be used to maintain excitation conditions for which with increasing pulse intensities the on-demand regime is reached. To provide a realistic picture of the achievable system performance, we include phonon-mediated processes in the theoretical caluclations. While preserving both high photon purity and indistinguishability, we find that although based on a higher-order emission process, for realistic system parameters on-demand Raman photon emission is indeed achievable with suitably tailored laser pulses.

\end{abstract}

%\keywords{}

\maketitle

\section{Introduction}

\begin{figure}[t]
 \includegraphics{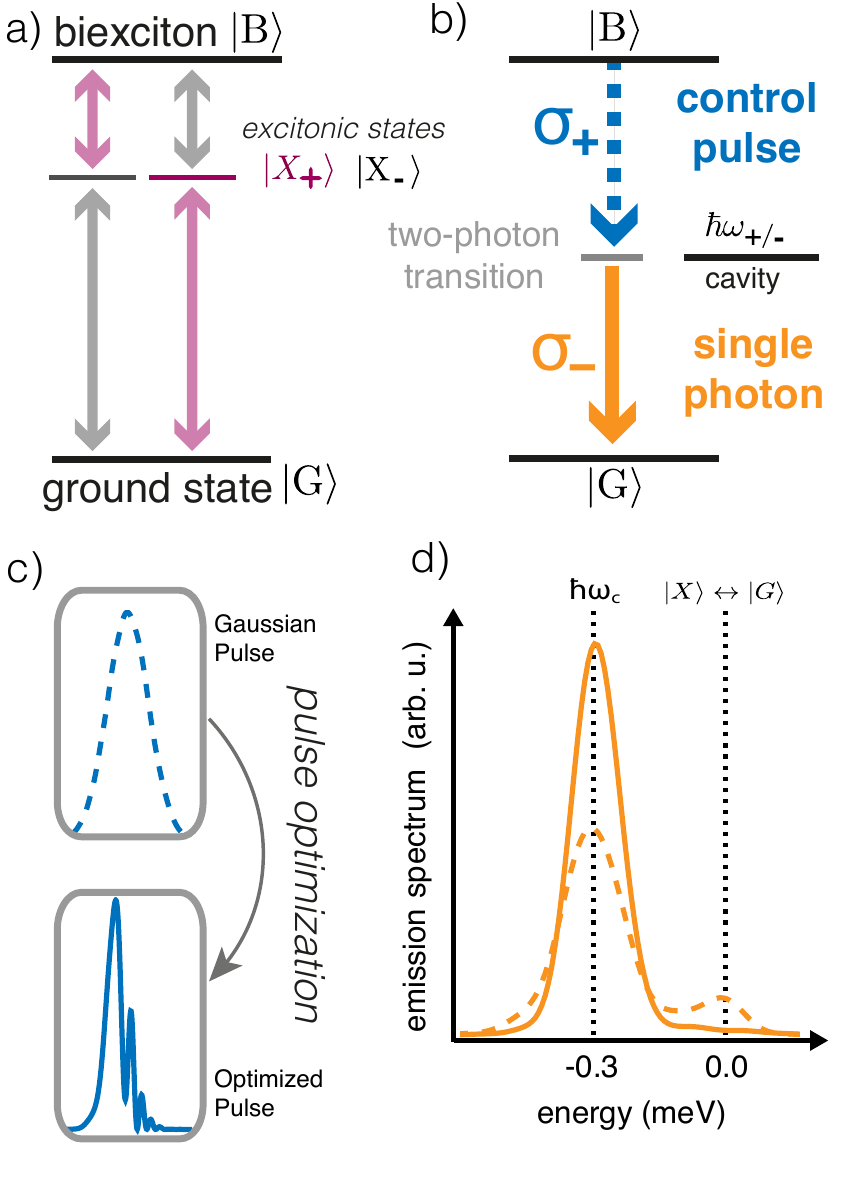}
 \caption{\label{fig:SPE:schematics:new}{a)}  {Sketch of the optical transitions in a semiconductor quantum dot in the circular polarization basis.} {b)} Sketch of the single photon Raman emission process from a direct two-photon transition between biexciton and ground state. A control pulse with $\sigma_+$ circular polarization is used to stimulate the system into a virtual state, from which a single photon with $\sigma_-$ circular polarization is then emitted as the system relaxes to its ground state. An optical cavity is used to enhance the single photon emission. In the present work, using a mask in Fourier-space, systematically optimized control pulses are used to maximize the single-photon emission probability. An example of the pulse optimization in shown in c). d) Sample emission spectrum with (solid line) and without (dashed line) control pulse optimization. Optimization in particular decreases the emission into competing decay channels through the biexciton exciton cascade.}
 \end{figure}

Single photon sources based on semiconductor quantum-dot systems are considered key components for integration into quantum computers \cite{Knill2001QuantumComputers} and quantum cryptographic applications \cite{Gisin2002QuantumCryptography}. 
These sources have to produce light with extraordinary quantum properties such as high indistinguishability, emission efficiency, and purity \cite{Ding2016OnDemandSinglePhotons,Hanschke2018SinglePhoton,Chen2018EntangledPhotonsEfficientExtractionAntenna,Huber2008StrainEntangledOnDemand,Liu2019EntangledEhotonBrightnessIndistinguishability,Wang2019OnDemandEntangledPhotonsFidelityEfficiencyIndistinguishability}. They are typically based on single photon transitions from one quantum dot state to another or the cascaded emission of photons from a higher electronic configuration in a quantum dot. Basic properties of the emitted photons such as polarization state and frequency are often predetermined by the chosen quantum-dot transitions and structure used and active control is difficult to achieve. Here we use a different approach and utilize a direct two-photon transition between the quantum-dot ground state and the biexciton state, which offers flexibility in the initial biexciton  state preparation \cite{stufler2006two,Michler2015RobustBiexcitonPreparation} with fidelities close to one and the emission of quantum light with up to two photons \cite{del2011generation,Ota2011TwoPhotonEmission,schumacher2012cavity}. Previously we demonstrated that the direct (non-linear) two-photon emission process from the biexciton can not only be used to emit a pair of photons but through a photonic Raman process also to emit a single photon with optical control \cite{heinze2015quantum,breddermann2016all}. In the latter case, as depicted in Fig.~\ref{fig:SPE:schematics:new}, a coherent control pulse drives the system from an occupied biexciton state into a virtual state inside the band gap from which the system then relaxes into its ground state by emitting a single photon \cite{heinze2015quantum}. This single-photon emission can be enhanced using an optical cavity. Related types of optical Raman processes which allow at least partial optical control of the properties of the emitted photon, such as polarization state and frequency, were previously studied in detail in different three-level systems \cite{Atature551,Santori2009,sweeney2014cavity,vora2015spin,heinze2015quantum,breddermann2018MicroscopicTheory}.

To unlock the full potential of quantum applications high single-photon emission probabilities are required, exceeding $66\%$ for linear quantum computing \cite{Varnava2008SourcesQuantumComputing} and $50\%$ for Boson sampling \cite{BosonSampling2017}. For the single photon emission of interest in the present work we have previously shown that the Raman resonance condition non-trivially depends on the shape of the control pulse triggering the emission \cite{breddermann2016all, breddermann2018MicroscopicTheory}. Moreover, the Raman process can also act as a source of excitation-induced quantum interference instead of as a source of single-photon emission \cite{breddermann2018MicroscopicTheory}. Both these aspects play an increasingly important role for elevated control pulse intensities and generally tend to undermine our quest for entering the on-demand regime for single Raman photon generation. 

In the present paper we demonstrate that a systematically optimized pulse can be used to steer the emission into the desired single photon channel and effectively suppress undesired emission and quantum paths in the system dynamics. In Fig.~\ref{fig:SPE:schematics:new} c) and d) we show a sample result for which a simple Gaussian pulse stimulates a single Raman photon emission into the cavity mode with energy $\hbar\omega_c$ (parameters of the Gaussian pulse where optimized to achieve maximum emission probability). Further optimization can then only be obtained using a non-trivial control pulse that does not only enhance the Raman single photon emission probability significantly but also reduces competing emission channels. On a more detailed level it is also important to note that the Raman single photon emission benefits from resonance enhancement when occurring spectrally close to the dipole-allowed quantum dot transitions. However, with the cavity mode nearby these transitions, cavity feeding assisted by longitudinal acoustic (LA) phonons also becomes important. We show that even including these additional processes, for realistic system parameters and using well-established and experimentally accessible methods of pulse shaping, the on-demand regime for emission of single Raman photons from a quantum-dot cavity system can indeed be reached while preserving important figures of merit such as indistinguishability and single photon purity.

\section{Theory}

In this section we lay out the theory used to describe the nonlinear excitation dynamics in the quantum dot cavity system considered. In Section \ref{ref:Hamiltonian} we introduce the system Hamiltonian which is used to calculate the system dynamics in Sec.~\ref{ref:polaron_master_equation} including couplings to the environment. We then summarize how to calculate the single photon emission spectra in Sec.~\ref{ref:Spectra} and photon indistinguishibility and purity in Sec.~\ref{ref:IndistPurity}.

\subsection{Hamiltonian} \label{ref:Hamiltonian}

We model the quantum dot by including the relevant electronic configurations which are the ground state $|G\rangle$, two excitons $|X_H\rangle$ and $|X_V\rangle$, and the biexciton state $|B\rangle$, with respective energies $E_G$, $E_i$, and $E_B$. We also include two degenerate cavity modes with frequencies $\omega_{H,V}$ and polarizations that correspond to those of the excitons. The free system Hamiltonian then reads:
\newcommand{\pro}[1]{\ensuremath{|#1\rangle\langle #1|}}
\newcommand{\proj}[2]{\ensuremath{|#1\rangle\langle #2|}}
\begin{equation}
\begin{split}
H_0 &= E_G \pro{G} +E_B\pro{B} \\  
        &+ \sum_{i=H,V} \big(\ {E_i \pro{X_i} + \hbar\omega_ib_i^\dag b_i} \big).\
\end{split}
\end{equation}
The electronic system couples to the cavity modes,
\begin{equation}
\Hqdc =  \sum_{i=H,V}{\Big(g\big(\proj{G}{X_i} b_i^\dag +\proj{X_i}{B}b_i^\dag\big) + \mathrm{h.c.}\Big)},
\end{equation}
with coupling strength $g$ and cavity photon operators $b_i^{(\dagger)}$. To trigger the single photon emission we include an off-resonant coherent laser pulse:
\begin{equation}
\Hqdl = \sum_{i=H,V}{\Big(\big(\proj{G}{X_i}\Omega_i^\ast(t) +\proj{X_i}{B}\Omega_i^\ast(t)\big) + \mathrm{h.c.}\Big)},
\end{equation}
with Rabi frequency $\Omega_i$. Here we assume zero fine-structure splitting for the excitons with $\delta_{fss}$. In this case we can simplify the following analysis by assuming circular polarization of the control laser. The desired single photon emission can then be detected in the other circular polarization channel as sketched in Fig.~\ref{fig:SPE:schematics:new}. In the circularly polarized frame the coherent light field and cavity photon operators take the following form:
\begin{equation}
\Omega_{\sigma^{\pm}}^* = \frac{1}{\sqrt{2}} (\Omega_H^* \pm \text{i} \Omega_V^*) \text{ and } b_{\sigma^{\pm}}^{\dagger} = \frac{1}{\sqrt{2}} (b_H^{\dagger} \pm \text{i} b_V^{\dagger}) 
\end{equation}
The exciton states in the circular basis are given by $\vert X_{\sigma^{\pm}} \rangle = \frac{1}{\sqrt{2}} (\vert X_H \rangle \pm \text{i}\vert X_V \rangle )$. Together the full Hamiltonian of the light matter interaction in the circular basis then reads
\begin{equation}
\begin{split}
H_{\mathrm{int}} &=  \Big(\proj{G}{X_{\sigma^{+}}} +\proj{X_{\sigma^{-}}}{B}\Big) \left( g b_{\sigma^{+}}^\dag +\Omega_{\sigma^{+}}^\ast(t) \right)+ \mathrm{h.c.}\\ 
& +\Big( \proj{G}{X_{\sigma^{-}}} +\proj{X_{\sigma^{+}}}{B}\Big) \left( g b_{\sigma^{-}}^\dag +\Omega_{\sigma^{-}}^\ast(t) \right)+ \mathrm{h.c.}
\end{split}
\label{H-circ}
\end{equation}
The relevant equations of motion in the polaron frame will be formulated in the following section.

\subsection{System Dynamics} \label{ref:polaron_master_equation}

The density matrix $\r$ of the quantum-dot cavity system in the polaron frame obeys the following Liouville-von Neuman equation \cite{heinze2015quantum,heinze2017polarization,manson2016polaron} 
%\textcolor{blue}{(solved in the interaction picture with respect to $H_0$!!!)} 
\begin{equation}
\begin{split}
  \dv{t} \r(t) &= \frac{1}{\i\h} \commutator{H'_S(t)}{\r} \\
                  &+  \mathcal{L}_\mathrm{cav}(\r) + \mathcal{L}_\mathrm{pure}(\r)  + \mathcal{L}_\mathrm{phonon}(\r) + \mathcal{L}'_\mathrm{rad}(\r) \label{eq:PolaronME},
  \end{split}
\end{equation}
which includes coupling to an environment via contributions of the type $\mathcal{L}_\mathrm{}(\r)$. The loss of cavity photons is included by
\begin{equation}
\mathcal{L}_{\mathrm{cavity}}(\rho_s) = \frac{\kappa}{2} \sum_{i=H,V}(2 b_i\rho_s b_i^{\dagger} -  b_i^{\dagger} b_i \rho_s - \rho_s  b_i^{\dagger} b_i )\, ,
\label{Cavity-Decay}
\end{equation}
where $\kappa$ is the cavity loss rate which is varied in fractions/multiples of $g$ to investigate both weak and strong coupling. We choose $g = \hbar/10 \text{ ps}^{-1}$, $g/\kappa \approx 0.35$ and a biexciton binding energy of $3 \text{ meV}$ if no other values are noted. To account for the decay of quantum-dot coherences a pure dephasing term
\begin{equation}
\mathcal{L}_{\mathrm{pure}}(\rho_s ) = -\frac{1}{2} \sum_{\chi , \chi' ; \chi \neq \chi'} \gamma_{\mathrm{pure}}^{\chi, \chi'} \vert \chi \rangle \langle \chi \vert \rho_s \vert \chi' \rangle \langle \chi' \vert\,
\label{Lindbladterm-Dephasing}
\end{equation}
with $\chi , \chi' \in \{ G, X_H,X_V, B \}$ is included. A linear increase of the pure dephasing rate $ \gamma_{\mathrm{pure}} $ at low temperatures $T$ is accounted for as $ \gamma_{\mathrm{pure}}(T)=  {}^{1\mu\text{eV}}/_K\, \cdot T$ \cite{laucht2009dephasing}.  Additionally, phonon-assisted optical transitions are included using a second order Born-Markov approximation  tracing out the phononic degrees of freedom.  The term obtained describing the electron-phonon interaction reads \cite{gustin2018pulsed, manson2016polaron}
\begin{equation}
\begin{split}
  &\mathcal{L}_\mathrm{phonon}(\r) = - \frac{1}{\h^2} \int_{0}^{\infty} \dd{\t} \\
  &\times \sum_{m = g, u} \qty( G_m(\t) \commutator{X_m(t)} {X_m(t-\t, t)\r(t)} + \hc
  )     \label{eq:phononLindblad} .
  \end{split}
\end{equation}
Here,  the polaron Green functions are given by \cite{roy2012polaron}
\begin{align}
  G_g(\t) &= \ev{B}^2 \qty( \cosh(\phi(\t)) - 1),  \label{eq:Gg} \\
  G_u(\t) &= \ev{B}^2 \sinh(\phi(\t)),  \label{eq:Gu}
\end{align}
with the phonon correlation function \cite{wilson2002quantum, heinze2017polarization}
\begin{equation}
  \phi(t) = \int_{0}^{\infty} \frac{J(\o)}{\o^2} \qty[ \coth(\frac{\h\o}{2\kbt} )\cos(\o\t) - \i \sin(\o\t) ]  \dd{\o}  \label{eq:phiTau}
\end{equation}
and the thermal average of the phonon bath displacement operator \cite{roy2011influence}
\begin{gather}
\ev{B} \equiv \ev{B_\pm} = \exp[- \frac{1}{2} \int_{0}^{\infty} \frac{J(\o)}{\o^2} \coth(\frac{\h\o}{2\kbt} ) \dd\omega ], \label{eq:evB} 
\end{gather}
where $T$ is the temperature of the QD sample. 
The main source of phonon-induced dephasing in InAs/GaAs QDs is a deformation potential induced by LA phonons \cite{Ramsay2010Damping,Krummheuer2002Dephasing,roy2011influence}. 
In this case, the phonon spectral function may be written as
\begin{equation}
  J(\o) = \a \o^3 e^{- \frac{\o^2}{2 \o_b^2} },  \label{eq:Jomega}
\end{equation}
where exciton-phonon coupling strength $\a$ and phonon cut-off frequency $\o_b$ are material parameters of the QD. 
Typical experimental values in InAs/GaAs QDs are $\a=\SI{0.03}{ps\squared} $ and $\h\o_\mathrm{b}=\SI{1}{meV} $ \cite{gustin2018pulsed}.  
The phonon--assisted operators are
\begin{equation}
  X_g = \mathcal{X}  + \hc \quad \text{and}\quad X_u = \i \qty(\mathcal{X}  - \hc), \label{eq:Xgu}  
\end{equation}
where
\begin{equation}
  \mathcal{X} = \sum_{j=\pol} 
  \Big(\pxjg + \pbxj\Big) \qty( g\bj + \Oj). \label{eq:mathcalX} 
\end{equation}
We note that in the numerical implementation all operators are treated in the interaction picture with respect to $H_0$ \cite{breddermann2016all}.

The optical quantum--dot transitions are time-dependent in the interaction picture $X_m(t-\t, t)$.  To calculate the $X_m(t-\t, t)$ exactly, we resort to solving the Heisenberg equation of motion \cite{gustin2018pulsed}
\begin{equation}
  \dv{\t} O(t-\t) = \frac{1}{\i\h}  \commutator{H'_\mathrm{I}(t-\t)}{O(t-\t)} + \pdv{\t} O(t-\t), \label{eq:heisenbergXTilde} \\
\end{equation}  
with $O(t-\t) \in$ $\{ \pxjg \bj, \pbxj \bj, \pxjg \O(t-\t), \pbxj \O(t-\t) \}$.
Integrating this differential equation from the initial condition $O(t - 0, t) = O(t) $ backwards until $t-\t$ yields $O(t-\t, t)$ and thus $X_m(t-\t, t)$ via \cref{eq:Xgu,eq:mathcalX}.
Note that in the interaction picture the involved operators carry an explicit time dependence, for instance 
\begin{equation}
  \pdv{\t} \qty[\pxjg \bj(t-\t)] = -\i (\oxj - \og - \omc) \pxjg\bj(t-\t). \label{eq:partDeri}
\end{equation}

We also consider radiative decay into modes other than the system cavity modes with  \cite{1367-2630-13-11-113014,roy2011influence,roy2012polaron}
\begin{equation}
\mathcal{L}_{\mathrm{rad}}(\rho_s ) = \gamma_{rad} \sum_{i=X_H, X_V}  \left( \mathcal{L}_{\vert G \rangle \langle i \vert } +\mathcal{L}_{\vert i \rangle \langle B \vert }  \right) (\rho_s)
\label{Leaky-Modes}
\end{equation}
where we chose $ \gamma_{rad}  =  2 \mu \text{eV}$ and
\begin{equation}
\mathcal{L}_{\mathrm{\sigma}}(\rho_s ) = (2 \sigma\rho_s \sigma^{\dagger} -  \sigma^{\dagger} \sigma \rho_s - \rho_s  \sigma^{\dagger} \sigma ).
\end{equation}
The polaron-transformed Hamiltonian $H'_S$ (\cref{eq:after_transform}) which appears in the  polaron master equation for the density matrix $\r$ scales the optical transitions in $H_S$ by a factor of $\ev{B}$. This rescaling also occurs in the radiative decay term with  $\mathcal{L}'_\mathrm{rad}(\r) = \ev{B}^2 \mathcal{L}_\mathrm{rad}(\r)$ \cite{roy2011influence, heinze2017polarization}.

\subsection{Single Photon Emission Spectrum}\label{ref:Spectra}

In Fig.~\ref{fig:SPE:schematics:new} d) we introduced the cavity emission spectrum, which is  known as the physical spectrum of light \cite{1367-2630-13-11-113014,Eberly:77,breddermann2016all}. It can be calculated  as
\begin{equation}
S_{i}(\mathcal{T},\omega) = \Re \int_0^{\mathcal{T}} dt \, \int_0^{\mathcal{T}-t}  d \tau \, \langle  b_i^{\dagger}(t) b_i(t+\tau) \rangle e^{i \omega \tau}
\label{Cavity-spectrum}
\end{equation}
for a given (circular) polarization $i$ up to a time $\mathcal{T}$ and it requires the evaluation of two-time expectation values which we calculate using the quantum regression theorem \cite{carmichael1999Methods}. We note, that in the cases considered here, the single photon emission process is completed in a time frame less than $60 \text{ ps}$ such that we can safely chose $\mathcal{T} = 60 \text{ ps}$ as a cut-off value for the time integrations.

\subsection{Indistinguishability and Purity} \label{ref:IndistPurity}

The purity and indistinguishability of an emitted  photon are crucial figures of merit of single photon sources. Those properties are essential for applications of quantum technology \cite{Senellart2017RevSPE}. The single photon purity quantifies whether a quantum light field contains more than one photon.  It is defined as the normalized equal-time two-photon expectation value \cite{fox2006QuantumOptics}
\begin{equation}
g^{(2)}_i(0) = \frac{ b_i^{\dagger}(t)  b_i^{\dagger}(t) b_i(t) b_i(t)}{\langle b_i^{\dagger}(t) b_i(t)\rangle^2}.
\end{equation}
The two-photon component of a pure single photon field equals zero.

The indistinguishability is of importance  whenever photon-photon interaction is vital \cite{Senellart2017RevSPE}. It can be measured in a Hong-Ou-Mandel interference experiment and reflects the joint detection probability at two photon detectors \cite{HOMexp1987}. We model this coincidence detection probability according to Ref. \onlinecite{gustin2018pulsed} as
\begin{equation}
p_c = \frac{\int_0^{\mathcal{T}} \int_0^{\mathcal{T}}  G_{HOM,i}^{(2)}(t,\tau) \text{ d}\tau \text{ d}t }{ \int_0^{\mathcal{T}} \int_0^{\mathcal{T}} \left(2  G_{pop,i}^{(2)}(t,\tau) - \vert \langle b_i(t+\tau) \rangle  \langle b_i^{\dagger}(t) \rangle \vert^2 \right) \text{ d}\tau \text{ d}t   }.
\end{equation}
Here 
\begin{equation} 
\begin{split}
G_{HOM,i}^{(2)}(t,\tau) &= \frac{1}{2} (  G_{pop,i}^{(2)}(t,\tau) + G_{i}^{(2)}(t,\tau) \\
&- \vert \langle b_i^{\dagger}(t+\tau) b_i(t) \rangle \vert^2 ),
\end{split}
\end{equation}
with $G_{pop,i}^{(2)}(t,\tau) =  \langle b_i^{\dagger} b_i \rangle(t) \langle b_i^{\dagger} b_i \rangle(t+\tau)$ and $G_{i}^{(2)}(t,\tau) =  \langle b_i^{\dagger}(t)b_i^{\dagger}(t+\tau) b_i(t+\tau) b_i(t) \rangle$ is the second order autocorrelation function.
The indistinguishability $\mathcal{I} =  1 - p_c$ has a maximal value of 1.

\section{Single Photon Emission Optimization} 

Single photon sources with a high emission probability or brightness are a desired key component for future quantum applications. Therefore, it is crucial to improve the single photon output of such sources.   In this paper we consider a single photon emission process based on the stimulated emission from a biexciton state in a quantum-dot, as outlined in the introduction. However, there also exist different competing spontaneous decay channels. Thus, it is necessary to isolate the contribution of the stimulated single photon emission  in the optimization process. To this end, in Sec.~\ref{Sec:RamanPop} below we introduce the single photon Raman emission probability which is then used to optimize the source's brightness. This optimization then results in an optimized optical control pulse that triggers the Raman photon emission process with its maximum yield. We parameterize the pulse in an experimentally accessible way based on well established pulse shaping technology as detailed in Sec.~\ref{Sec:PulseShaping} with additional remarks concerning the optimization in Sec.~\ref{Sec:Opt}.

\subsection{Raman Population} \label{Sec:RamanPop}

We are interested in the photon emission of the quantum light source. The total photon emission probability from the optical cavity with photon polarization $i$ during a time span $\mathcal{T}$ is given by \cite{pathak2010coherently}
\begin{equation} 
\mathcal{P}_{ems,i}(\mathcal{T}) = \kappa \int_0^{\mathcal{T}} N_i(t) \text{ d} t,
\end{equation}
where $\kappa$ is the cavity decay rate and $N(t) = \text{ tr}(\rho(t)\, b_i^{\dagger} b_i)$ is the photon population. This expression combines all photons from different decay/emission events. We are, however, only interested in the emission of a certain photon from the optically controlled, direct transition from the biexciton to the ground state. Following Ref.~\onlinecite{breddermann2018MicroscopicTheory} we introduce the single photon Raman emission  probability of polarization $i$ as
\begin{equation} 
\mathcal{P}_{\mathcal{R}, i}(\mathcal{T}) = \kappa \int_0^{\mathcal{T}} N_{\mathcal{R}, i}(t) \text{ d} t.
\label{RamanEmissionProb}
\end{equation}
Here, we are interested in the circularly polarized {population} of the Raman photon  $N_{\mathcal{R}, \sigma_-}(t)$. This is obtained from the Heisenberg equation for the mean of the circularly polarized cavity photon number operator up to second order in the hierarchy with the Raman process 
\begin{equation} 
 \mathcal{R}_{\sigma_-}(t) = \langle  \vert G \rangle \langle B \vert b_{\sigma-}^{\dagger}(t) \Omega_{\sigma_+}^{*}(t)  \rangle.
\end{equation}
Integrating the second order equation while keeping only the terms proportional to the Raman process yields
\begin{equation} 
\begin{split}
 N_{\mathcal{R}, \sigma_-}(t) &=  \frac{2 g \langle B \rangle^2 }{\hbar^2} \text{ Re } \int _0^t \int_0^{t'} e^{-\kappa(t-t')} \\
 &\times \left( e^{W_{BX}(t')-W_{BX}(t'')} -e^{W_{XG}(t')-W_{XG}(t'')}   \right) \\
 &\times \mathcal{R}_{\sigma_-}(t'') \text{ d}t'' \text{ d}t',
\end{split}
\end{equation}
where we have $W_{\alpha \beta}(t) = - \text{i}  \int _0^t  \omega_{\alpha \beta, c}(t') \text{ d}t' $, and $\omega_{\alpha \beta, c}(t) := \omega_{\alpha}-\omega_{\beta}-\omega_{c} -\frac{\text{i}}{2} (\kappa + \bar{\Gamma_{\alpha \beta}}(t))$, with $\omega_{\alpha / \beta}$ relating to the quantum-dot energies and  $\hbar \omega_{c}$ is the energy of the cavity mode. The quantum-dot decay terms read
\begin{eqnarray} 
\bar{\Gamma}_{X_{\sigma-}G}(t) = \Gamma_{XG} + \Gamma_{L,+}^{X_{\sigma+}G}(t) + \Gamma_{L,+}^{BX_{\sigma-}}(t), \\
\bar{\Gamma}_{BX_{\sigma+}}(t) = \Gamma_{BX} + \Gamma_{L,-}^{GX_{\sigma+}}(t) + \Gamma_{L,-}^{X_{\sigma-}B}(t).
\end{eqnarray}
The pure dephasing and radiative decay are included via  $\Gamma_{XG} = \gamma_{\text{pure}} + \gamma_{\text{rad}}$ and $\Gamma_{BX} = \gamma_{\text{pure}} + 3\gamma_{\text{rad}}$ \cite{breddermann2016all} while the phonon-mediated processes driven by the laser field $\Gamma_{L,\pm}^{\alpha/\beta}$ are approximated by the analytical rates from Refs.~\onlinecite{manson2016polaron,Kumar2017}. We neglect the cavity photon mediated rates as $\ev{B}^2 g^2 \ll \kappa$ in our case. When we consider the single photon emission without the phonon-assisted processes we set $\ev{B}=1$ and $\bar{\Gamma}_{\alpha \beta}(t) = \Gamma_{\alpha \beta}$.

\subsection{Pulse Shaping} \label{Sec:PulseShaping} 

We model the pulse shaping required for the numerical optimization according to the output of a $4-f$ pulse shaper \cite{Sussman20084fpulseShaper}. Here, an input beam is focused onto a spatial light modulator (SLM) which makes it possible to introduce a frequency dependent phase. The SLM applies a mask $M$ to  an input field $\Omega^{\text{in}}$ in frequency domain
\begin{equation}
\Omega^{\text{SLM}} (\omega) = M(\omega) \Omega^{\text{in}} (\omega) 
\end{equation}
with $M(\omega) = A_M(\omega) e^{i \phi_M(\omega)}$.
A common choice is to choose $A_M(\omega) = 1$ which implies that the pulse intensity will be conserved by the SLM \cite{Debnath2012ChirpedLaserExcitonPhonon}. 
The phase mask is given by
\begin{equation}
\phi_M(\omega) = \alpha \cos [2 \pi \gamma(\omega -\omega_L) + \delta ]  + \eta (\omega -\omega_L)^2.
\end{equation}
The first term models a periodic phase commonly used in optimal quantum control \cite{mathew2015simultaneous, mathew2015simultaneous}. The second term is a quadratic phase which introduces a linear chirp in the time domain \cite{Malinovsky2001ARP} and which is used to account for pulse induced shifts during the single photon emission process. 
\textcolor{black}{These shifts are a major contribution to the overall pulse shape and can be analytically understood in the context of Raman emission from three-level systems (cf. footnote Ref.~45 of Ref.~\onlinecite{breddermann2018MicroscopicTheory}).}
We choose a Gaussian shaped input field 
\begin{equation}
\Omega^{in}_{\sigma +} (t) = \hbar \sqrt{ \frac{\mathcal{E} _0 \pi}{4 \text{ ps } \sigma}} e^{- ( \frac{t-t_0}{\sigma} )^2 } e^{i \omega_L t} \, \text{ and }\, \Omega^{in}_{\sigma -} (t) = 0
\end{equation}
 with  pulse width $\sigma$, center $t_0$, dimensionless measure of energy $\mathcal{E}_0$ and $\hbar  \omega_L = E_B -E_G -\hbar\omega_C +\Delta_L$, where $\Delta_L$ is a small detuning accounting for pulse induced resonance shifts. 
 The pulse shaping method conserves the pulse energy, hence the dimensionless amplitude $\mathcal{E}_0$ acts as a constraint for the optimized pulse.

\subsection{Optimizing the Single Photon Emission}\label{Sec:Opt}

We aim to optimize the single photon emission as quantified by the Raman emission probability $\mathcal{P}_{\mathcal{R}, \sigma-}$ in equation (\ref{RamanEmissionProb}). The parameterization of the pulse introduces a $7$-dimensional optimization problem with parameters $ x = \{\alpha, \gamma, \delta, \eta, \sigma, t_0, \Delta_L \}$. In this space we numerically seek \cite{LinNonlinOPT}
\begin{equation}
\max_{x} \, \mathcal{P}_{\mathcal{R}, \sigma-}(x).
\end{equation}
We restrict the set $x$ to parameters suitable for common SLMs and chirped pulses based on Refs.~\onlinecite{Glassl2013BiexStatePrep,mathew2011optimal} with
\begin{align*}\label{SLM-Bounds}
0 \le \alpha \le 2\pi, &\qquad 0 \text{ meV}^{-1} \le \gamma \le 2 \text{ meV}^{-1},\\
0 \le \delta \le 2\pi,& \qquad -25 \text{ meV}^{-2} \le \eta \le 25 \text{ meV}^{-2},\\
15 \text{ ps} \le t_0 \le  30 \text{ ps}, &\qquad 2.5  \text{ ps} \le \sigma \le 5 \text{ ps}\,.\\
\end{align*}
The numerical optimization of the single photon Raman emission probability is performed by applying advanced nonlinear optimization algorithms as implemented in the IPOPT (Interior Point OPTimizer) software package \cite{wachter2006ImplementationInteriorpoint,TutorialIPOPT,TrustRegionIP}.  
This requires the numerical evaluation of both the gradient and the Hessian matrix of the circular Raman emission. 
To calculate the occurring partial derivatives with machine precision we employ \emph{algorithmic differentiation} (AD) \cite{bartholomew-biggs2000,adolc,GriewankAD,CheckpointingADOLC}.  
The \emph{CoDiPack}  library \cite{sagebaum2019HighPerformanceDerivative} is used for AD as it shows a high performance for the problem considered in this paper. 
We employ forward mode AD which is more efficient than reverse mode for the given number of optimization parameters and parallelize the computations of the partial derivatives.

\section{Results \& Discussion}

In this section we discuss the main results of the single photon Raman emission optimization. We begin with Sec.~\ref{SPRE} where we study the Raman output depending on the cavity detuning in the weak optical coupling regime. In Sec.~\ref{SPEaP} we examine the potential for on-demand operation for higher quality cavities approaching the strong coupling regime, while preserving other important photon properties such as indistinguishability and purity. 

\subsection{Single Photon Raman Emission}\label{SPRE}

\begin{figure}[t]
 \includegraphics[scale=0.52]{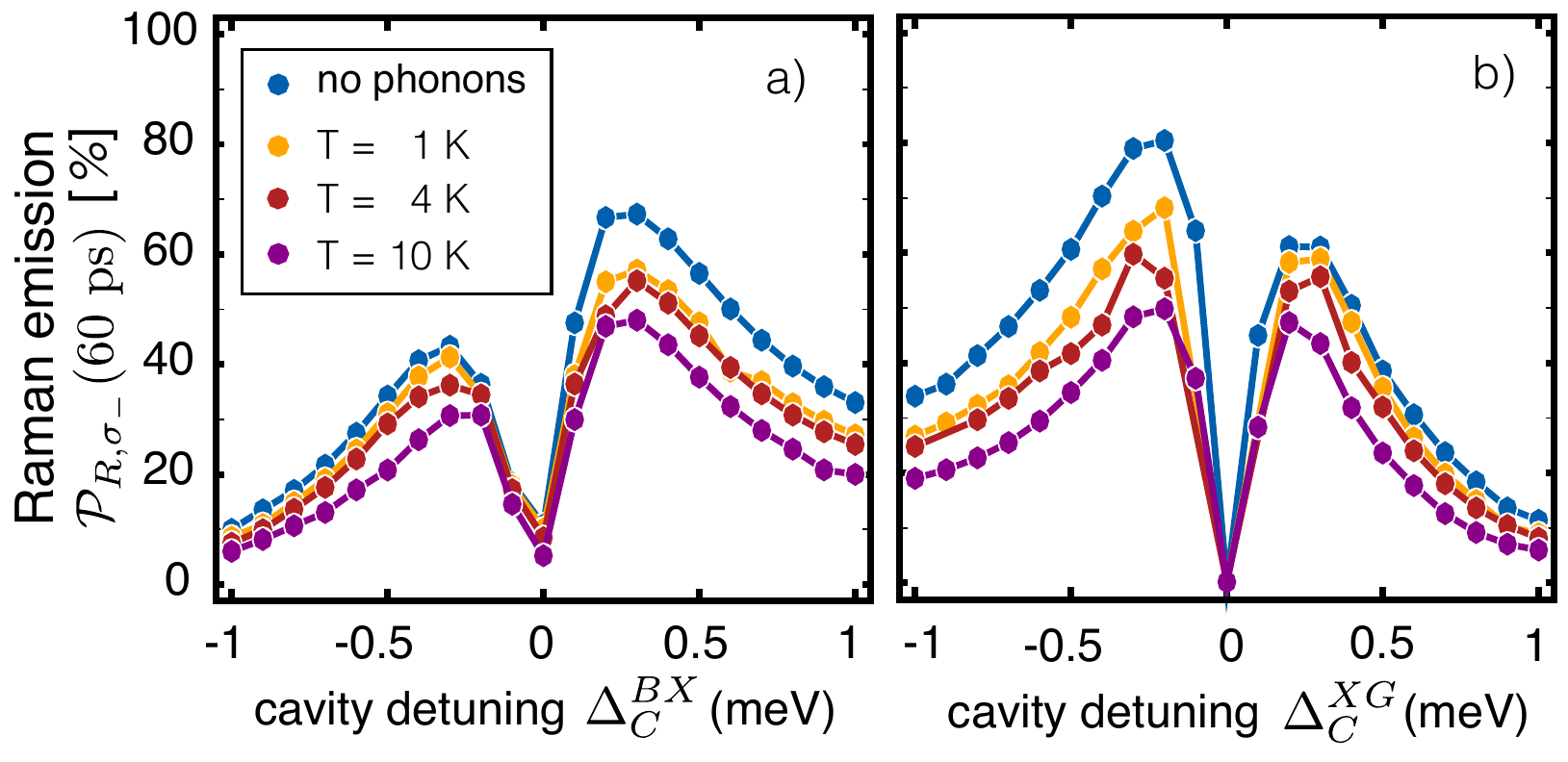}
 \caption{\label{fig:SPE:results:PulseDetuning}  Dependence of the single photon Raman emission probability on the detuning of the cavity from the biexciton to exciton transition $\Delta_C^{BX}=\hbar \omega_C -E_B + E_X$ in a) and the exciton to ground state transition $\Delta_C^{XG}=\hbar \omega_C - E_X$ in b).  Results are shown for zero phonon coupling as well as for calculations including the phonon-assisted processes at temperatures  $T = 1\, \text{K}$,  $T = 4\, \text{K}$, and $T = 10\, \text{K}$ for \textcolor{black}{$g/\kappa=0.35$}. } 
 \end{figure}

We start by studying the dependence of the single photon Raman emission depending on the detuning of the cavity mode from the closest nearby quantum--dot resonance. We assume that at time $t=0$ the quantum dot system is initialized in the biexciton state and no photons are present in the optical cavities. Fig.~\ref{fig:SPE:results:PulseDetuning} depicts the resulting Raman emission probability for optimized pulses that are constrained to the dimensionless pulse amplitude $\mathcal{E}_0 = 4$. Results are shown with phonon mediated transitions at $T = 1 \text{ K}$, $T = 4 \text{ K}$ and $T = 10 \text{ K}$ as well as without phonon coupling for comparison. In the cases considered, the single photon emission process is completed for times smaller than $60 \text{ps}$. Spectrally approaching the single-photon resonances in the quantum dot system leads to a resonance enhancement also in the nonlinear system response and enhances the emission probability of the single Raman photons. Exactly on resonance, however, at $\Delta_C^{BX / XG}=0$, where one would expect the total cavity emission probability to peak (not shown in Fig. \ref{fig:SPE:results:PulseDetuning}), the Raman emission rate is close to zero due to quantum interference of different emission channels \cite{breddermann2018MicroscopicTheory}. As a consequence we find the resonance enhancement from nearby real transitions to be strongest if the cavity is positioned between the electronic quantum dot resonances, with a slight asymmetry favoring the exciton to ground state transition. The strongest single Raman photon emission is found for detunings $\Delta_C^{BX/XG} = \pm 0.3 \text{meV}$. We note that in order to realistically benchmark the system dynamics, this close to the resonance condition phonon-mediated processes must be taken into account. Overall, we observe that the coupling to phonons reduces the single photon emission probability. The phonon-assisted processes open alternative and competing decay channels and in addition $\ev{B} < 1$ directly reduces the generated Raman photon number average. At low temperatures the phonon bath can more easily absorb than emit quanta of energy. Thus, phonon-assisted absorption and emission prefer opposite spectral detunings \cite{roy2011influence}. Phonon-assisted emission of a photon into the cavity mode occurs if $\Delta_C^{BX/XG} < 0$. In these cases, tuning the cavity to the biexciton to exciton resonance does only increase the background emission. If, however, the cavity is tuned to the exciton to ground state transition phonon-assisted transitions hinder the control laser to stimulate the higher order Raman process. Consequently, the Raman emission probability is decreased.  On the other hand, if $\Delta_C^{BX/XG} > 0$ these effects are weak, but now the control laser is negatively detuned from the closest quantum-dot resonance so that the laser may induce optical transitions via the phonon side bands. However, in the case of $\Delta_C^{XG} > 0$, the phonon interaction mainly introduces another decay path to the system. We find an overall low degree of second order coherence of the emitted photons with $g^{(2)}_{\sigma-}(0)$ being well below $0.1$.

We note that for a medium-quality cavity with \textcolor{black}{$g/\kappa=0.35$ or $Q=7447$ at $880 \text{ nm}$} we find that the Raman emission probability at low temperatures ($T = 4$ K) can reach \textcolor{black}{$60$ $\%$} and as such is already sufficient for quantum computing applications such as Boson sampling \cite{BosonSampling2017}.

\subsection{Single Photon Emission and Properties}\label{SPEaP}

\begin{figure}[t]
 \includegraphics[scale=0.37]{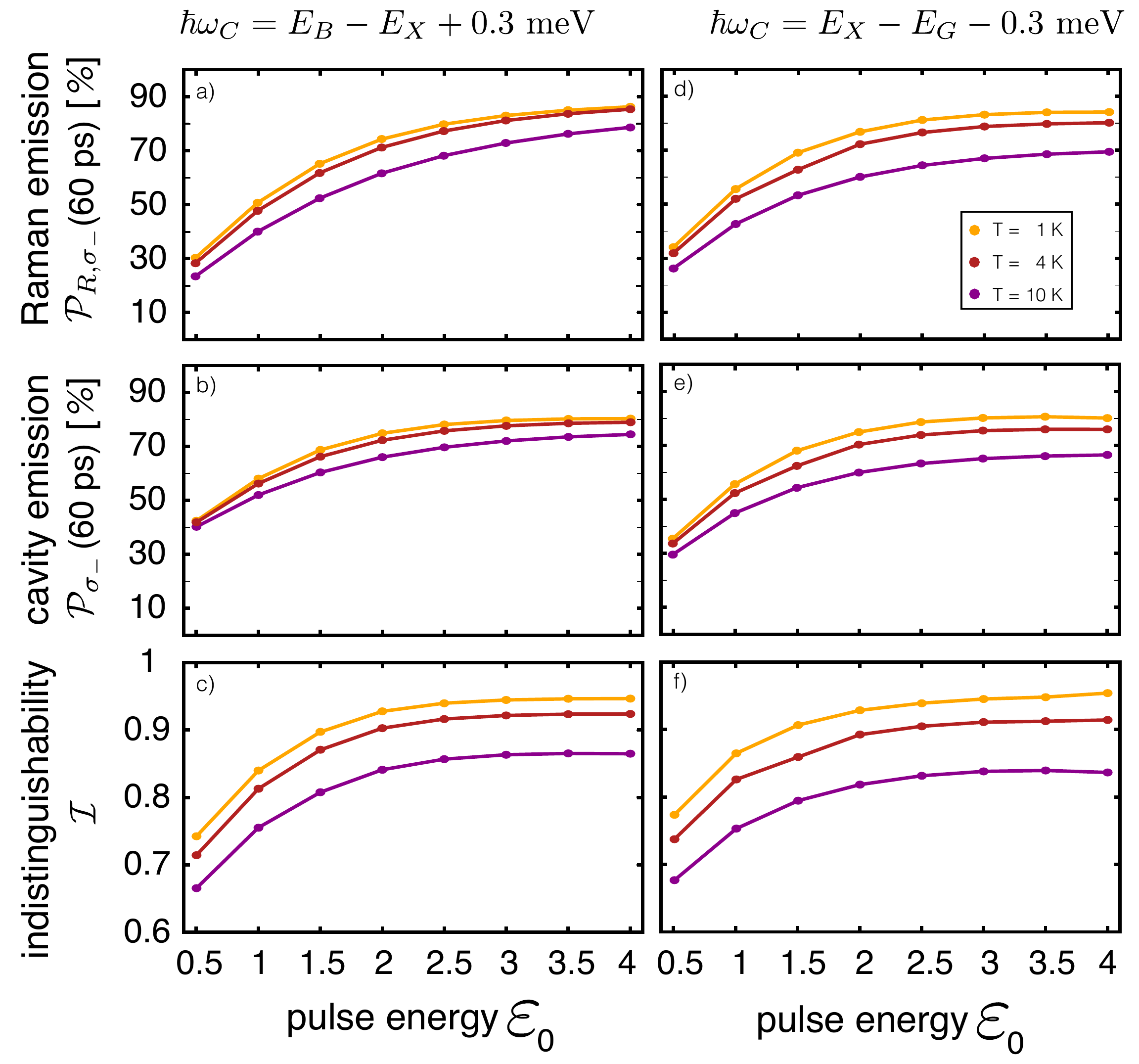}
 \caption{\label{fig:SPE:results:Indistinguishability} Dependence of  the single photon emission probability and indistinguishability on the control pulse strength. The influence of electron-phonon coupling is shown for the single photon Raman emission probability (a,d), the total cavity emission (b,e), and the indistinguishability (c,f) depending on the pulse strength for the two most promising cavity detunings from Fig.~\ref{fig:SPE:results:PulseDetuning}. Here, a high--Q cavity with $g = \kappa$ at $Q \approx 21000$ is used.} 
 \end{figure}

Above we only considered a medium-quality cavity which did not reach the required threshold of a photon emission probability of $2/3$ for on-demand emission \cite{Varnava2008SourcesQuantumComputing}. In the present chapter we further increase the light-matter coupling in our analysis. Now, a high-Q cavity with $g = \kappa$ is used. In Fig.~\ref{fig:SPE:results:Indistinguishability} we show the total photon emission probability of the $\sigma_-$-polarized cavity mode, the Raman emission probability and the indistinguishability of the photons after the optimization process. One sees that a single photon emission probability can be reached that is now high enough to be considered on-demand. We analyze the two situations most suitable for single photon emission with $\Delta_C^{BX}= +0.3\text{ meV}$ and $\Delta_C^{XG}= -0.3\text{ meV}$. Again, it is assumed that at time $t=0$ the quantum dot system is initialized in the biexciton state and no photons are present in the optical cavity modes. With increasing pulse energy $\mathcal{E}_0$ the Raman emission and the total cavity emission rise. The higher light-matter coupling partly compensates for the detrimental influence of the electron-phonon coupling. In general, as the pulse energy increases the phonon-assisted effects become more pronounced, since they scale with the square of the pulse energy as given in equation (\ref{eq:phononLindblad}). This leads to a saturation in achievable photon output at large pulse energies. We find a maximum Raman emission probability of $\mathcal{P}_{R,\sigma_-} \approx 85\, \%$ in the case of $\Delta_C^{BX}= +0.3\text{ meV}$ and a slightly lower value of $\mathcal{P}_{R,\sigma_-} \approx 80\, \%$ for $\Delta_C^{XG}= -0.3\text{ meV}$ at the typical experimental temperature of $T = 4 \text{ K}$ and $\mathcal{E}_0=4$. The difference $\mathcal{P}_{R,\sigma_-}  - \mathcal{P}_{\sigma_-} $, which can be  associated with destructive quantum interference \cite{breddermann2018MicroscopicTheory}, is marginal at high pulse energies, so that the quantum light source can be characterized as an on-demand source with emission probabilities of $\sim 80\, \%$. At higher temperatures of $T = 10 \, \text{ K}$, the optimized emission probability still exceeds the on-demand limit for $\Delta_C^{BX}= +0.3\text{ meV}$. However, in the case of $\Delta_C^{XG}= -0.3\text{ meV}$ the emission probability is about $10 \, \%$ lower because, even in the case of a positively detuned laser from the biexciton to exciton  transition, with increasing temperature the phonon bath is more likely to emit quanta of energy to bridge the energy gap $\Delta_C^{XG}$ for pulse induced optical transitions.

In Fig.~\ref{fig:SPE:results:Indistinguishability} (c,f) we depict the calculated indistinguishabilities for both detunings. At first, the indistinguishability rises with increasing pulse amplitude $\mathcal{E}_0$ until it plateaus because of phonon-induced coupling. At higher temperatures, the indistinguishability is reduced for $\Delta_C^{XG}= -0.3\text{ meV}$ in comparison with $\Delta_C^{BX}= +0.3\text{ meV}$. Here, starting with an occupied biexciton the $\sigma_+$-polarized pulse can populate the $\vert X_-\rangle$ exciton with rising temperature more easily resulting in the emission of $\sigma_-$-polarized photons into the cavity mode, which consequently reduces the indistinguishability of the triggered $\sigma_-$-polarized Raman photon. However, at low temperatures the phonon-mediated process is suppressed allowing high values of $\mathcal{I} \approx 0.95$ at $T = 1 \text{ K}$ and $\mathcal{I} \approx 0.92$ at $T = 4 \text{ K}$. Additionally, since the off-resonant high-Q cavity suppresses the (phonon-assisted) cascaded two-photon emission, an overall high single photon purity is found. While we observe the highest purity at the lowest temperature, we find a low degree of second order coherence $g^{(2)}_{\sigma-}(0) < 0.01$ even at $T = 10  \text{ K}$.

\begin{figure}[t]
 \includegraphics[scale = 0.35]{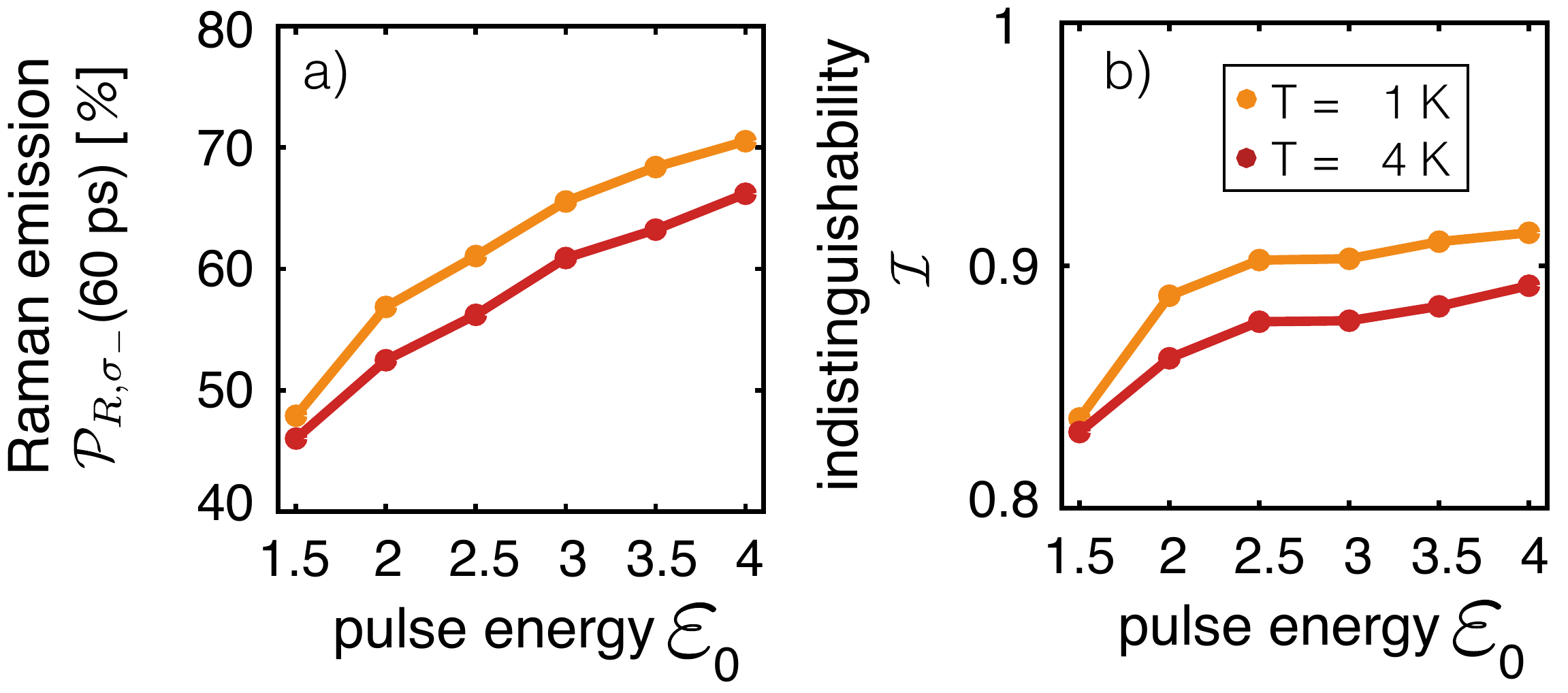}
 \caption{\label{fig:SPE:results:BiexInit}  Single photon emission with optical biexciton initialization. A two-photon absorption process is used to prepare the biexciton state from the initial ground state $\rho(0)=\pro{G}$ with a peak occupation of $N_B\approx0.72$. Depicted are the single photon Raman emission probability a) and the indistinguishability b) depending on the strength of the control pulse for the case of $\Delta_C^{BX}= +0.3\text{ meV}$. Again, a high--Q cavity with $g = \kappa$ and $Q \approx 21000$ is used. }
\end{figure}

Lastly, we analyze the effect of the fidelity of the initially prepared biexciton state. 
So far we have considered a maximum initial fidelity of $N_B(t_I=0) = 1$ where $t_I\ge0$ is the time when the maximum biexciton occupation is reached. 
To simulate the case of $N_B(t_i)<1$ we use a Gaussian pulse to incompletely drive the system from the ground state ($\rho(0)=\pro{G}$)  into the biexciton state by two-photon absorption with a peak occupation of $N_B\approx0.72$. 
In Fig.~\ref{fig:SPE:results:BiexInit} we depict the Raman emission probability and the indistinguishability for $\Delta_C^{BX}= +0.3\text{ meV}$.
%with \textcolor{magenta}{dashed lines}.  
It should be noted that the photon emission probabilities can exceed $N_B$ because the decaying biexciton does emit two photons. 
The photon emission probabilities as well as the indistinguishability are reduced for all values of the pulse amplitude and temperature.   
The single photon purity decreases only slightly with the pulse amplitude as $g^{(2)}_{\sigma-}(0) < 0.002$. 

Although we find that almost all of the initial biexciton occupation can be used for the creation of a single Raman photon through the control pulse optimization, the quantum efficiency of the single photon source is limited by the fidelity of the initial biexciton occupation. We note that the latter can be optimized separately. However, even with significantly reduced initial biexciton occupation we find on-demand single Raman photon emission from the biexciton within reach at low temperatures and sufficiently high pulse energies.

\section{Conclusion}

We have studied the systematic optimization of a Raman single photon source based on a partially stimulated two-photon emission process from the quantum dot biexciton inside an optical cavity. We find that the underlying Raman process is most efficient if the optical cavity used to enhance the light-matter coupling is placed near and between the electronic transitions of the quantum dot. We find that non-trivial shapes of the pulses triggering the Raman photon emission are key to suppress competing emission mechanisms. For an initially fully populated biexciton state and a sufficiently high-Q cavity (here with $Q=21000$ and $g=\kappa$), we numerically demonstrate single Raman photon emission with a probability of $\sim 80 \%$ and simultaneously high indistinguishability of $\sim 92 \%$ and high single photon purity with $g^{(2)}(0) < 0.01$. Even with incomplete optical biexciton initialization the on-demand regime can still be reached for realistic system parameters.

\begin{acknowledgments}
This work was supported by the Deutsche Forschungsgemeinschaft (DFG) through the collaborative research center TRR142 (grant No. 231447078, project A03) and Heisenberg program (grant No. 270619725) and by the Paderborn Center for Parallel Computing, PC$^2$.
\end{acknowledgments}

\appendix

\section{Interaction With LA Phonons}

\subsection{QD Hamiltonian Including Electron-Phonon Coupling} 
\label{sub:qd_hamiltonian_with_phonon_coupling}

In the case of InAs/GaAs QDs driven near resonance, the interaction with phonons is predominantly governed by the coupling to LA phonons \cite{roy2011influence}. By modeling the phonon bath as harmonic oscillators with wave vector $\vec{q}$ and energy $ \h\o_{\vec{q}} $ the Hamiltonian of the entire system may be written as \cite{roy2012polaron,Hargart2016SimDressing} 
\begin{equation}
  H = H_0 + H_S + H_\mathrm{B} + \Hqdp \\
\end{equation}
where  $H_S = \Hqdc + \Hqdl$ is the interaction Hamiltonian of the QD--cavity system.
The Hamiltonian of the phonon bath is given by
\begin{equation}
   H_\mathrm{B} = \sum_{\vq} \h\o_{\vq}  \acq \aq,
\end{equation}
while
\begin{equation}
  \Hqdp = \sum_{\mathclap{s = \bi,\xh,\xv}} \dyad{s}{s} \sum_{\vq} \h\l_{\vq}^i \qty( \acq + \aq  )
\end{equation}
describes the electron-phonon interaction.
Here, $\acq$ and $\aq$ are bosonic creation and annihilation operators for a phonon in the mode $ \vq $. 
The exciton-phonon coupling strength can be quantified by real constants $ \l^i_{\vq} $ \cite{gustin2018pulsed, manson2016polaron}.
Furthermore, for an ideal QD we have $  \l^{\bi}_{\vq} = 2\l^{\xh}_{\vq} = 2\l^{\xv}_{\vq} $ \cite{hohenester2007phonon}.

\subsection{Polaron Transformation}

At this point, a commonly used approach \cite{manson2016polaron, roy2012polaron} is to apply the transformation
\begin{equation}
  H' = e^{S} H e^{-S} \qq{with} S = \sum_{\mathclap{s=\bi,\xh,\xv}} \dyad{s}{s}  \sum_{\vq} \frac{\l_{\vq}^s}{\oq} \qty( \acq - \aq) . \label{eq:polaron_transform} 
\end{equation}
This transforms the Hamiltonian into the {polaron frame},  removing the explicit appearance of the electron-phonon interaction $\Hqdp$ \cite{wilson2002quantum,heinze2017polarization}.
Using the Baker-Campbell-Hausdorff formula the transformation defined by \cref{eq:polaron_transform} may be carried out analytically \cite{gustin2018pulsed}, yielding  
\begin{equation}
  H' = H_0' + H_S' + H_\mathrm{B} + H_\mathrm{I} \label{eq:after_transform}.
\end{equation}
Here, the transformed Hamiltonian of the QD and the cavity modes is given by
\begin{equation}
    H_0' = \sum_{\mathclap{s=\levels}} E_s' \dyad{s}{s} + \sum_{\mathclap{i=\pol}} \h\omc \bcj \bj   \label{eq:transformedH0}
\end{equation}
where $E_s'=E_s - \sum_{\vq} \nicefrac{\l_{\vq}^s}{\o_{\vq} }  $. 
This {polaron shift} of the QD energy levels can be included in the original definition of $E_s$ and therefore be disregarded \cite{manson2016polaron}. 
The transformed QD interaction Hamiltonian
\begin{equation}
  H_S' = \ev{B} H_S \label{eq:H_Sstrich}
\end{equation}
is scaled by the thermal average of the phonon bath displacement operator \cite{roy2011influence}.
Since $\ev{B} < 1$, this effectively decreases the coupling strengths $g$ and $\O(t)$.

Lastly, the new interaction Hamiltonian of the phonon--assisted optical transitions is given by 
\begin{equation}
  H_I = \z_g X_g - \z_u X_u \label{eq:Qdph_transformed}
\end{equation}
with fluctuation operators $\z_g  = \frac{1}{2} (B_+ + B_- - 2 \ev{B} )$ and $\z_u = B_+ + B_-$ \cite{heinze2017polarization}.

\section{Optimized Pulses}
\label{appendix:OptimizedPulses}

Table \ref{tab:optparameters} lists sets of parameters for which an optimal emission was  found as shown in Fig. \ref{fig:SPE:results:Indistinguishability}  and Fig. \ref{fig:SPE:results:BiexInit} for $\mathcal{E}_0 = 3$ at different temperatures. 
Either an initially prepared biexciton state is assumed (system initially in state $\vert B\rangle$ with zero photons) or the biexciton state is prepared by two-photon absorption from the ground state of the quantum dot  (system initially in state $\vert G\rangle$ with zero photons). \textcolor{black}{We note that an initialization pulse may also introduce pulse-dependent spectral shifts, which are reflected in the optimized parameters of the (temporally overlapping) control pulse.}

\begin{table}[ht]
    
     \caption{ List of parameters of those pulses for which optimal emissions is found as shown in Fig.~\ref{fig:SPE:results:Indistinguishability} (initial condition $\vert B\rangle$) and Fig.~\ref{fig:SPE:results:BiexInit} (initial condition $\vert G\rangle$) for $\mathcal{E}_0 = 3$ at different temperatures. }
    \label{tab:optparameters}
    
    \begin{tabular}{ p{2.5cm}||p{1.25cm}|p{1.25cm}||p{1.25cm}|p{1.25cm}  }
         \hline
         \hline 
         \centering
         initial condition & $\vert G\rangle$ &$\vert G \rangle$ &$\vert B \rangle$ &$\vert B \rangle$\\ [1ex]
        \centering
         T [K] & 1 & 4 & 1 & 4\\[0.75ex] 
         \hline
         \hline
         \centering
         $\Delta_L$ [meV] & 0.17 &0.17  &0.16 &0.15\\ [1ex]
          \hline
          \centering
         $\alpha$ [rad] & 2.05 & 3.14 & 3.14 & 3.14\\ [1ex]
          \hline
          \centering
         $\gamma$ [meV$^{-1}$] &1.31 &1.14 &1.13 &1.17 \\ [1ex]
         \hline
         \centering
         $\delta$ [rad] & 2.69 &2.72 &-0.74 &-0.74\\ [1ex]
          \hline
          \centering
         $\eta$ [meV$^{-2}$] & 12.94 &0.80 &-25.0 & -24.44 \\ [1ex]
          \hline
          \centering
         $t_0$ [ps] & 26.75 &25.56 &22.38 &22.73\\ [1ex]
          \hline
          \centering
         $\sigma$ [ps] & 4.97 & 5.0 & 5.0 &  4.99\\  [1ex]
         \hline
         \hline
         \centering
         pulse area [$\pi$] & 8.2 & 8.2 & 8.1 &  8.2\\  [1ex]
         \hline
         \hline
    \end{tabular}
\end{table}

\bibliography{Literatur}

\end{document}